\documentclass[aps,jcp,twocolumn,groupedaddress]{revtex4}
\usepackage{graphicx,color,subfigure}
\usepackage[draft]{pdfcomment}
\begin{document}
\title{Nonlinear mechanics of thermoreversibly associating dendrimer glasses}
\author{Arvind Srikanth}
\author{Robert S. Hoy}
\email{rshoy@usf.edu}
\affiliation{Department of Physics, University of South Florida, Tampa, FL, 33620}
\author{Berend C. Rinderspacher}
\author{Jan W. Andzelm}
\affiliation{U.S. Army Research Laboratory, Weapons and Materials Research Directorate, Aberdeen Proving Ground, Aberdeen, MD 21005}
\date{\today}

\begin{abstract}
We model the mechanics of associating trivalent dendrimer network glasses with a focus on their energy dissipation properties.  Various combinations of sticky bond (SB) strength and kinetics are employed.  The toughness (work-to-fracture) of these systems displays a surprising deformation-protocol dependence; different association parameters optimize different properties.  In particular, ``strong, slow'' SBs optimize strength, while ``weak, fast'' SBs optimize ductility via self-healing during deformation.  We relate these observations to breaking, reformation, and partner-switching of SBs during deformation.  These studies point the way to creating associating-polymer network glasses with tailorable mechanical properties.
\end{abstract}

\maketitle

\section{Introduction}
Structural materials are used in a variety of applications with
widely ranging, often complex mechanical and chemical requirements. 
For example, a car's crumple zone must retain its shape
and stability during normal operation while irreversibly deforming during a collision.
Several aspects determine the suitability of a material for a given
application: for example, its ductility, strength, and toughness.
All of these tend to depend in nontrivial ways upon deformation conditions such as strain protocol and strain rate, with corresponding differences in maximal energy dissipation and thermal response.
This diversity of requirements produces a need for materials with tunable mechanical properties. 

Incorporation of supramolecular complexes into polymeric systems provides a route to address
these demands. 
The integration of thermoreversibly associating groups into polymers
produces a wide variety of complex behavior arising from the finite
lifetime of the ``sticky'', thermoreversible bonds.\cite{brunsveld01,rotello08,annable93,binder07,tanaka02,seiffert12}
Examples of thermoreversible bonds include $\pi$-$\pi$-stacking,\cite{burattini10,burattini11,colquhoun02,sivakova05} hydrogen bonding,\cite{cordier08,Edwards2013,colquhoun12,vanbeek07,hirschberg99} and metal ligand bonding.\cite{burnworth08,burnworth11,yount03,xu11,el-ghayoury03,schmatloch02,schmatloch03,fustin07,kumpfer10,doi:10.1021/ma401077d}  
In associating polymer systems (APs), the presence of thermoreversible bonds leads to exquisitely
tunable rheological properties, particularly when the product
$\dot{\epsilon}\tau_{sb}$ of the strain rate $\dot{\epsilon}$ and
sticky bond lifetime $\tau_{sb}$ is of order unity.\cite{yount03,yount05,loveless05,hoy09} 
AP networks are also capable of dramatic ``self-healing'' under a variety of conditions, e.g., after fracture.\cite{cordier08,colquhoun12,burnworth11}

Such properties have led to an explosion of interest in these systems' rheology over the past decade, in systems ranging from dilute solutions to dense melts.  
Most studies to date have focused on melts well above their glass transition temperature $T_g$.  
However, glassy associating polymer systems are also of great interest for their potential as energy-dissipating materials.  
Potential benefits include rate and temperature dependencies that are stronger than found in non-associating glassy systems, and thermoplastic-elastomer-like response when the sticky bonds form a percolating network (similar to that formed by the chemical crosslinks in thermoplastics).  This response can be achieved in systems that are easily melt-processable when the sticky bonds are sufficiently weak that no such percolating network is present at processing temperatures above $T_g$.
Complex, nontrivial plastic flow and fracture behavior can arise from the fact that sticky bonds  behave like transient covalent bonds and produce a correspondingly transient entanglement network.  The latter is of especial interest due to its potential to produce self-healing materials with enhanced fracture toughness arising from recombination of bonds that have broken during deformation, a phenomenon with no analogue in non-associating polymer systems.

Analytic and quasi-analytic approaches to AP dynamics and mechanics, e.g.\ Refs.\ \cite{leibler91,tanaka92,rubinstein98,rubinstein01,tanaka02,semenov06,indei07,semenov07}, have made many useful, experimentally verifiable predictions, including nonlinear behaviors such as shear thickening and strain hardening.\cite{pellens04,pellens04b,tripathi07}
However, for the sake of tractability, theories have generally neglected one or more features of AP systems that are likely essential to capturing their behavior under certain ambient conditions.
For example, as temperature drops towards $T_g$, attractive, non-associative interactions, such as van der Waals forces between non-sticky monomers, become increasingly important.\cite{baschnagel05}
For such systems, molecular simulations are necessary to capture the essential features.

In this paper we use hybrid molecular dynamics/Monte Carlo simulations to examine the glassy mechanical response of systems composed of model trivalent dendrimers.
Such molecules are of interest as simple, small-molecule building blocks for AP glasses\cite{cordier08} that offer advantages in terms of processability, reversibility, and functionality.  
We show that these systems exhibit a complex mechanical response
wherein the sticky bond thermodynamics, temperature $T$, chemical kinetics, and strain rate are all relevant.  We examine deformation through fracture under two deformation modes commonly employed in experiments - uniaxial stress and uniaxial strain - and show that different sticky bonding parameters optimize different quantities such as strength, ductility, work-to-fracture and the tendency for ``self-healing'' during deformation.

\begin{figure}
\includegraphics[width=2.5in]{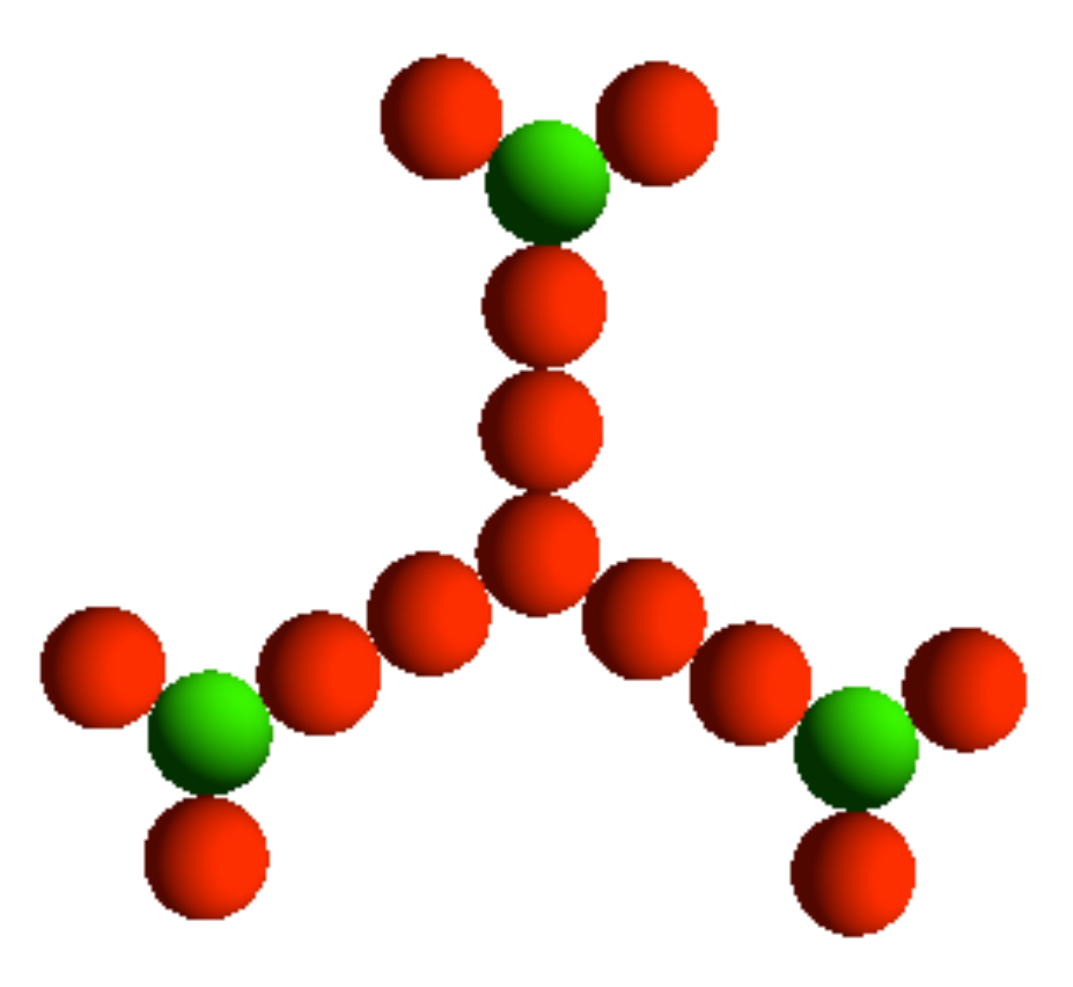}
\caption{Coarse-grained model of trivalent dendrimers; green monomers represent the sticky groups.  The ``floppy ears'' serve as a qualitative, coarse-grained representation of volume-excluding moieties.}
\label{fig:molecules}
\end{figure}

\section{Model and Methods}

We employ the hybrid molecular dynamics/Monte Carlo (MD/MC)
algorithm described at length in Ref.~\cite{hoy09} as well as further below.
A generic trivalent model monomer is shown in Fig.~\ref{fig:molecules}.
The red spheres in the model represent unreactive, linking
components. They have no angle potentials and merely serve to exclude
volume and define the general topology. They therefore do not
represent any chemically distinct species, but rather any chemical
configuration that realizes the coarse topology and is sufficiently
flexible.
The green spheres in Fig.~\ref{fig:molecules} represent reactive or
associative components that can bond with
each other thermoreversibly.
These associating, ``sticky'' monomers (SM), again, do not represent any
specific chemistry.
The terminating red spheres serve as volume excluding moieties that
provide proper separation of the SM.
Realizations of such a system may be $CH((CH_2)_nL(CH_2)_x)_3$ or
$N((CH_2)_nL(CH_2)_x)_3$ dendrimers with appropriately large $n$ and
$x$ \cite{cordier08,Edwards2013}.
Here, for the purposes of modeling generic properties, the size of the SMs and the mobility of ``open'' (unbonded) SMs are equal to that of regular monomers.
We employ the ``Y'' architecture as the simplest possible model of a dendrimer; although the arms are far too short to be entangled, the overall mechanical response of the glassy systems described below indicates that the response is entangled-like and thus not (to first order) architecture-specific.

All monomers have mass $m$ and interact via the truncated and shifted Lennard-Jones (LJ) potential $U_{LJ}(r) = 4u_{0}[(a/r)^{12} - (a/r)^{6} - (a/r_{c})^{12} + (a/r_c)^{6}]$, where $r_{c}=2^{7/6}a$ is the cutoff radius and $U_{LJ}(r) = 0$ for $r > r_{c}$.
Covalent bonds between adjacent monomers on a chain are modeled using the
finitely extensible nonlinear elastic potential $U_{FENE}(r) = -(1/2)(kR_{0}^2) {\rm ln}(1 - (r/R_{0})^{2})$, with the canonical\cite{kremer90} parameter choices $R_{0} = 1.5a$ and $k = 30u_{0}/a^{2}$.   
In this study, following the majority of bead-spring studies on permanently crosslinked systems (e.~g.~Refs.\ \cite{combinedgrest90,svaneborg08}), we employ flexible chains with no angular potential.
We express all quantities in units of the LJ bead diameter $a$, intermonomer energy $u_{0}$, and the LJ time $\tau_{LJ} = \sqrt{ma^{2}/u_{0}}$. 

\begin{figure}
\includegraphics[width=3.375in]{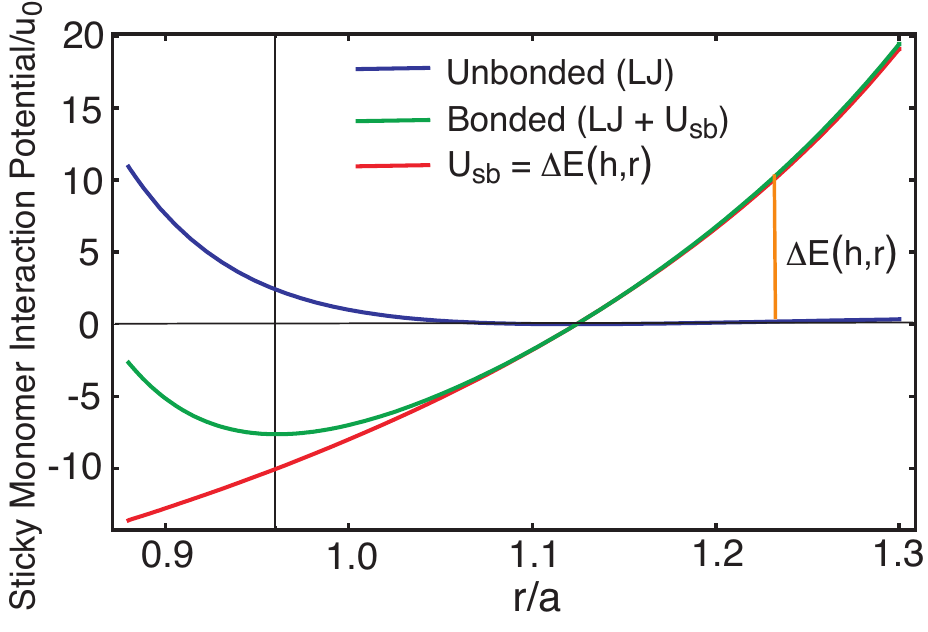}
\caption{Interaction potential for sticky monomers as a function of bond length.  The blue line represents the Lennard-Jones term, the red line represents $U_{sb}(h,r)$ for $h = 10u_0$, and the green line represents the total bonded potential for a ``closed'' SB.  The orange vertical line represents the $r$-dependent energy change for SB association.}
\label{fig:interactionpot}
\end{figure}

Sticky bonds (SBs) interact via the same potential employed in Ref.\ \cite{hoy09}, i.e.
\begin{equation}
\begin{array}{cccc}
U_{sb}(r) & = & U_{FENE}(r) - h, & r < R_0\\
& & & \\
& = & 0, & r > R_0
\end{array}
\label{eq:ufenesb}
\end{equation}
where $h$ is the thermodynamic strength of the sticky bonds
(Fig.~\ref{fig:interactionpot}).
Newton's equations of motion are integrated using MD with a time step $\delta t = .008\tau_{LJ}$, and sticky bonds are formed and broken using standard Metropolis Monte Carlo\cite{hoy09}.  
MC moves are executed every 25 MD timesteps, i.e.~every $\tau_0 = 0.2\tau_{LJ}$.
A typical SB pair is considered for dissociation (or an SM pair for association) once every $\tau_{MC}$. 
As in Ref.~\cite{hoy09}, the use of different $\tau_{MC}$ corresponds in a qualitative, coarse-grained manner to modeling ligands $L$ with different chemical kinetics; for example, bulky ligands would have slower kinetics.\cite{yount03,yount05,loveless05}
At each MC timestep, a fraction $f_{MC}$ of open SB pairs are considered for SB formation and the same fraction of closed SB pairs are considered for dissociation.
The characteristic kinetic time\cite{hoy09} is $\tau_{MC} = \tau_0/f_{MC}$, e.g.\ for $f_{MC} = .002$, $\tau_{MC} = 100\tau_{LJ}$.
SM association is binary; SMs are bonded to either zero or one other SM at any given time.
Thus the model has second-order association (i.e. $2A \to A_2$) and first-order dissociation kinetics ($A_2 \to 2A$), where $A$ represents a sticky monomer.

We prepare our samples as follows. All systems have $N_{ch} = 16000$
chains of the architecture shown in Fig.~\ref{fig:molecules} and are prepared at a monomer density $\rho
= .85a^{-3}$.  Periodic boundary conditions are applied along all
three directions of simulation cells that are initially cubic with side lengths $L_x=L_y=L_z=L_0$. 
We thoroughly equilibrate
systems at $T=T_{eq}=1.125$ and zero pressure for several times the
sticky bond lifetime $\tau_{sb}(h,\tau_{MC},T)$, verifying that the
fraction of ``closed'' SBs as a function of equilibration time $t$, $p_c(t)$, has plateaued according to fits of $p_c(t)$ to the function\cite{hoy09}
\begin{equation}
p_c(t) = d-\frac{\left(d^2-1\right) \tanh \left(2z\sqrt{d^2-1} t 
   \right) + d\sqrt{d^2-1}}{d \tanh \left(2z\sqrt{d^2-1} t \right)+\sqrt{d^2-1}},
\label{eq:specpoft}
\end{equation}   
i.e.\ $p_c^{eq} \simeq d(h)$ where $z = \rho c_{st} k_f(h)$, $d = 1 + k_b(h)/4z$, $c_{st}$ is the SM concentration (i.e.\ the mole fraction: $c_{st}=3/16$), and $k_f(h)$ and $k_b(h)$ are respectively the rate constants for SM association and SB dissociation.
We then perform a slow, zero-pressure quench (rate $\dot{T} = -10^{-5}/\tau_{LJ}$) to a final temperature $T_f < T_g \simeq 0.43$.  Temperature is controlled using a Langevin thermostat.
At such temperatures, more than 99.5\% of all SMs are bonded into SBs.

We focus on the mechanical properties of systems deep in the glassy state, i.e.\ $T_f$ well below $T_g$.
After quenching, we perform mechanical properties tests using two
standard deformation protocols.\cite{rottler03,rottler03b}  Uniaxial
stress and uniaxial strain runs are performed at $T_f=0.3\approx 0.7 T_g$ at two constant tensile strain rates, $\dot{\epsilon} = \dot{L}_z/L_{0} = 10^{-5.5}/\tau_{LJ}$ and $\dot{\epsilon} = \dot{L}_z/L_{0} = 10^{-4}/\tau_{LJ}$.  In the uniaxial stress simulations, pressure along the transverse directions is maintained at zero using a Nose-Hoover barostat, and deformation takes place at nearly constant volume.  We thus report stress-strain curves against the Green-Lagrange strain $g(\lambda) = \lambda^2 - 1/\lambda$, where $\lambda=L_z/L_0$; systems having response analogous to linear rubber elasticity produce stress-strain curves linear in $g(\lambda)$, while chain-stretching produces supralinear behavior (``Langevin'' hardening\cite{treloar75}) and SB-breaking produces sublinear behavior. 
Uniaxial strain simulations are performed at constant cross-sectional area ($L_x = L_y = L_0$), and stress-strain curves are reported vs. $\lambda$.

As described below, we will examine the mechanical properties of associating dendrimer glasses for a variety of $h$, $\tau_{MC}$ (Table \ref{tab:systems}), and deformation protocols and show nontrivial dependencies on each.  
In particular, we will show that fracture toughness and ductility cannot both be optimized by a single choice of $h$ and $\tau_{MC}$ for different deformation protocols, and relate this to protocol-dependent self-healing.
During deformation, we monitor stress $\sigma$ (reported in units of $u_0/a^{3}$), the total mechanical work performed up to strain $\epsilon=ln(\lambda)$ is
\begin{equation}
W(\lambda) = \int_{0}^{\lambda} \sigma d\lambda',
\label{eq:work}
\end{equation}
and several metrics of SB formation and recombination.
$P_{surv}(\epsilon)$ is the fraction of initially closed sticky bonds that persist continuously from zero strain through strain $\epsilon$.
$P_{recomb}(\epsilon)$ is the fraction of SM pairs $A-B$ (initially bonded at $\epsilon = 0$) that have  broken and recombined by strain $\epsilon$.
Finally, $P_{switch}(\epsilon)$ is the fraction of $A-B$ pairs (initially bonded at $\epsilon = 0$) that have performed a ``partner switch'' (i.e.\ formed a bond $A-C$, where $C \neq B$) by strain $\epsilon$.
As we will show, $P_{surv}$, $P_{recomb}$, and $P_{switch}$ are important metrics for understanding fracture and self-healing.

\begin{table}[htbp]
\caption{Systems employed in this study.  ``W'' stands for ``weak'' and ``S'' for ``strong'' sticky bonds, ``f'' for ``fast'' and ``s'' for ``slow'' kinetics, and the numerals at the end indicate applied strain rates.}
\begin{ruledtabular}
\begin{tabular}{lcccl}
System Name & $h$ & $\tau_{MC}$ & $\dot{\epsilon}$
& \parbox[t]{8em}{Line convention in plots.}\\
Wf4 & 10 & 1 & $10^{-4}$ & red, solid, thick\\
Ws4 & 10 & 100 & $10^{-4}$ & red, dashed, thick\\
Sf4 & 15 & 1 & $10^{-4}$ & blue, solid, thick\\
Ss4 & 15 & 100 & $10^{-4}$ & blue, dashed, thick\\
Wf55 & 10 & 1 & $10^{-5.5}$ & red, solid, thin\\
Ws55 & 10 & 100 & $10^{-5.5}$ & red, dashed, thin\\
Sf55 & 15 & 1 & $10^{-5.5}$ & blue, solid, thin\\
Ss55 & 15 & 100 & $10^{-5.5}$ & blue, dashed, thin\\
\end{tabular}
\end{ruledtabular}
\label{tab:systems}
\end{table}

\section{Results}

\begin{figure}
\includegraphics[width=3.25in]{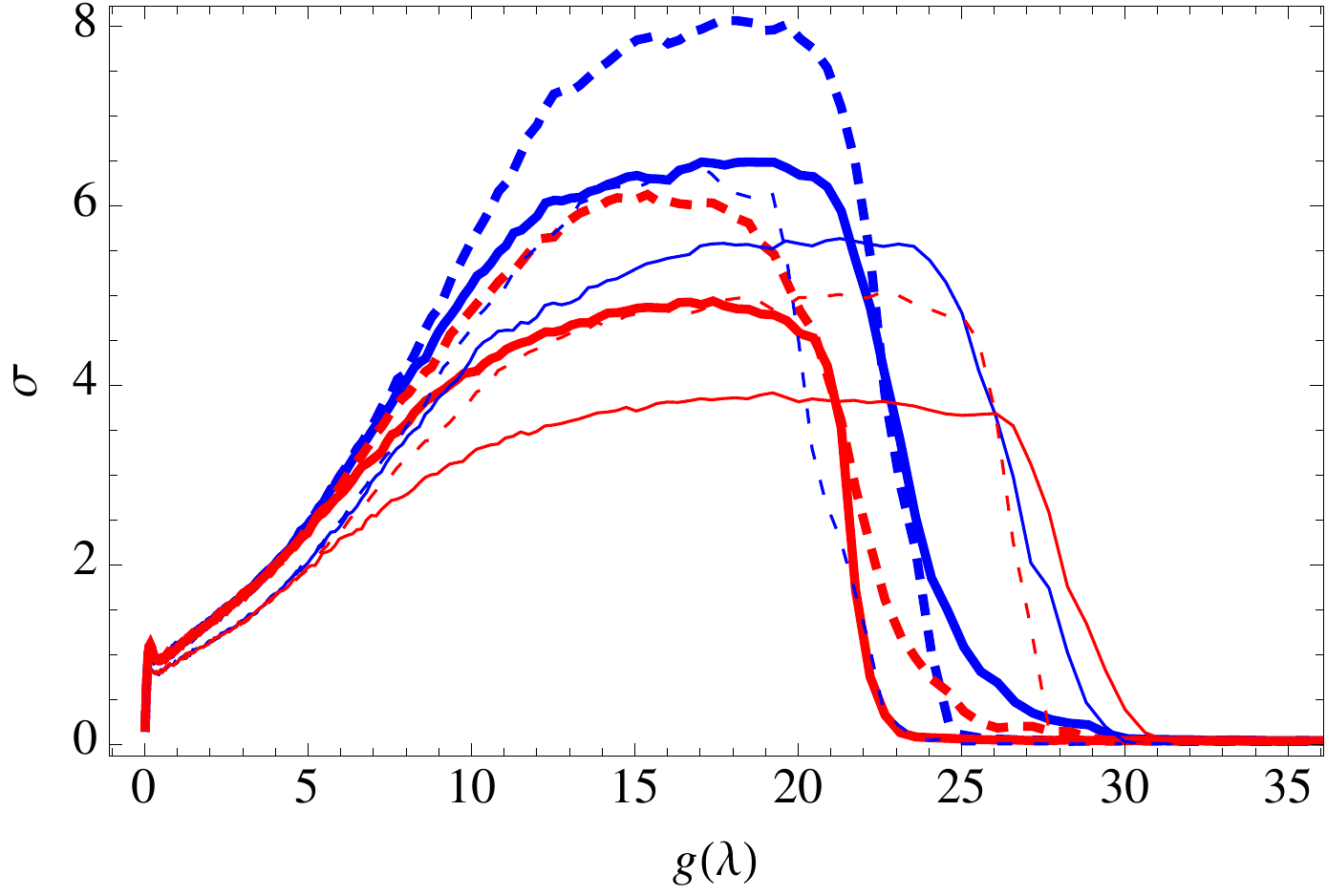}
\includegraphics[width=3.25in]{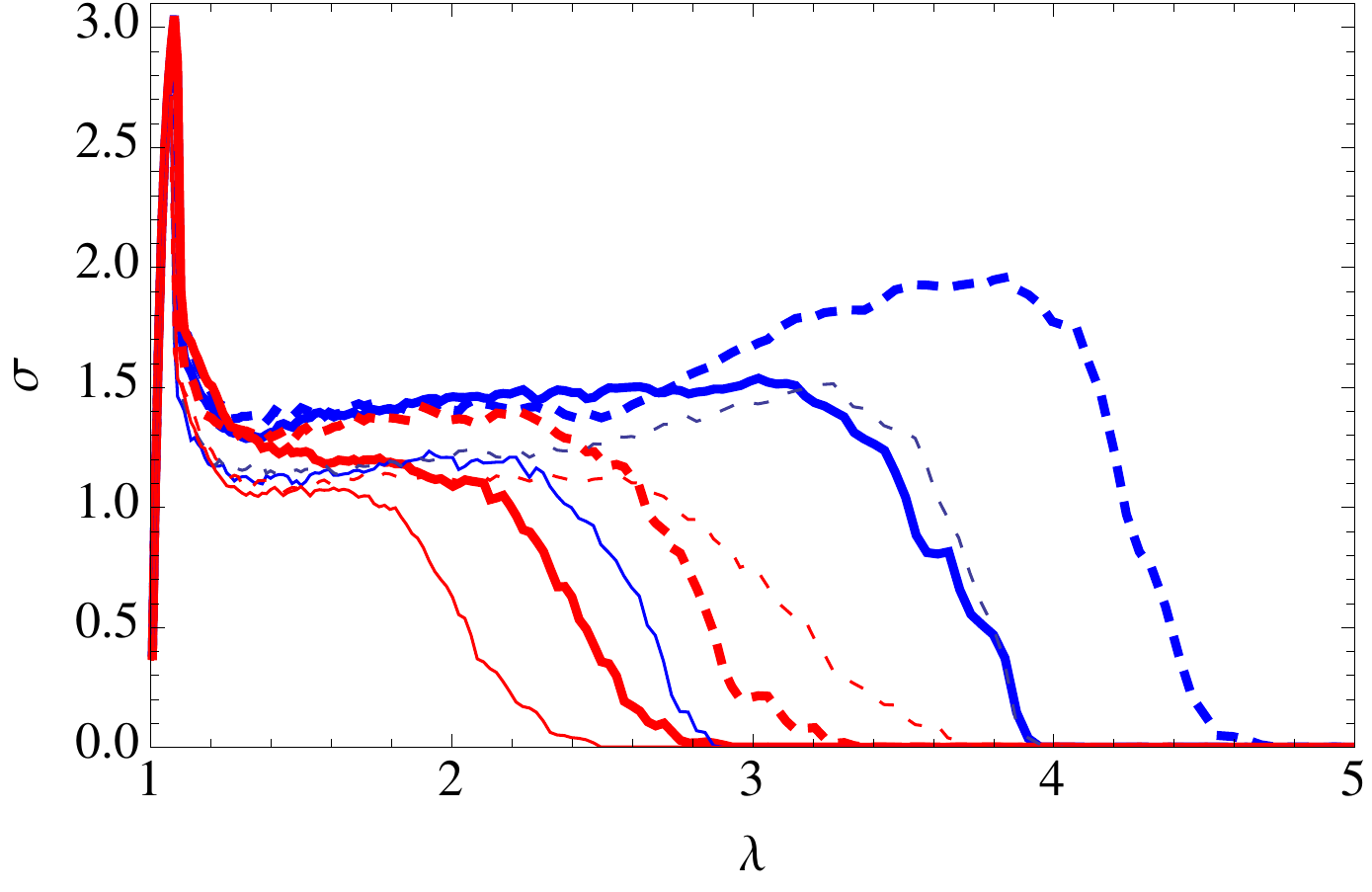}
\caption{Mechanical responses of studied systems deformed at $T = 0.3$. Solid and dashed lines  indicate results for $\tau_{MC} = \tau_{LJ}$ and $\tau_{MC} = 100\tau_{LJ}$, respectively. Blue and red curves indicate results for $h=15$ and $h=10$, respectively. Thick and thin lines show results for $\dot{\epsilon}=10^{-4}$ and $\dot{\epsilon}=10^{-5.5}$, respectively.
Top: uniaxial tensile stress vs. Green-Lagrange strain, $g(\lambda) = \lambda^{2}-1/\lambda$; Bottom: uniaxial tensile strain deformation, stress vs. stretch $\lambda = L_z/L_z^{0}$. }
\label{fig:stressstrain2}
\end{figure}

In this section, we analyze the effects of SB thermodynamics, kinetics, and deformation rate on the nonlinear mechanics of thermoreversibly associating dendrimer glasses.  
Figure \ref{fig:stressstrain2} shows results for uniaxial stress and strain deformation for all systems considered in this study.  
Note that for both protocols, although our systems possess no physical
entanglements, systems show a response typical of thermoplastic
elastomers or entangled glasses\cite{haward97}; initial yield followed by strain hardening (for uniaxial stress) or craze drawing (for uniaxial strain).
Results for systems without sticky monomers (not shown) show far lower stresses, indicating SBs are the dominant contributor to the toughness.  This is unsurprising since our systems have no topological entanglements, but serves to illustrate the dramatic influence of SBs on glassy mechanical response.

\begin{figure}
\includegraphics[width=3in]{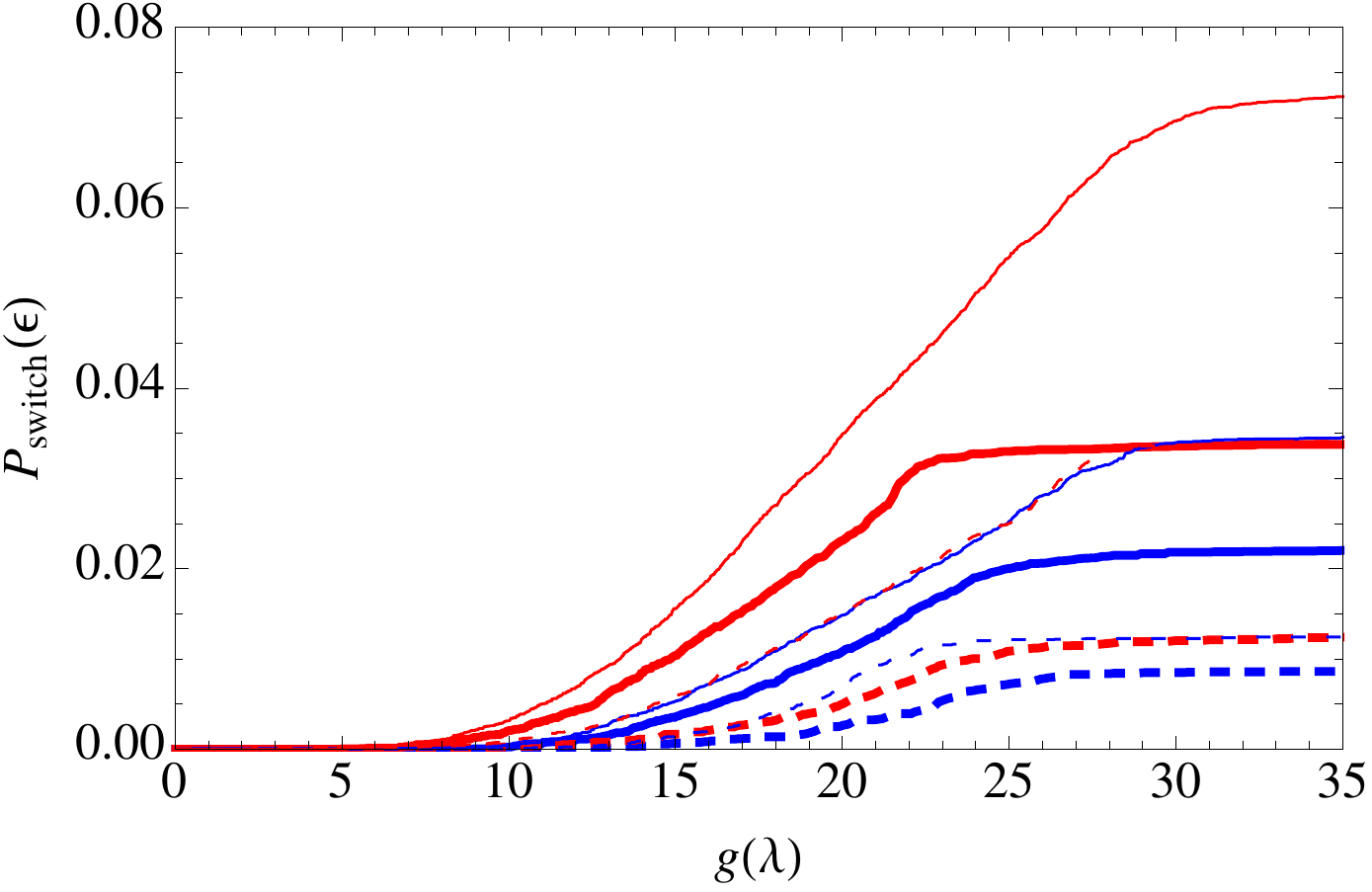}
\includegraphics[width=3in]{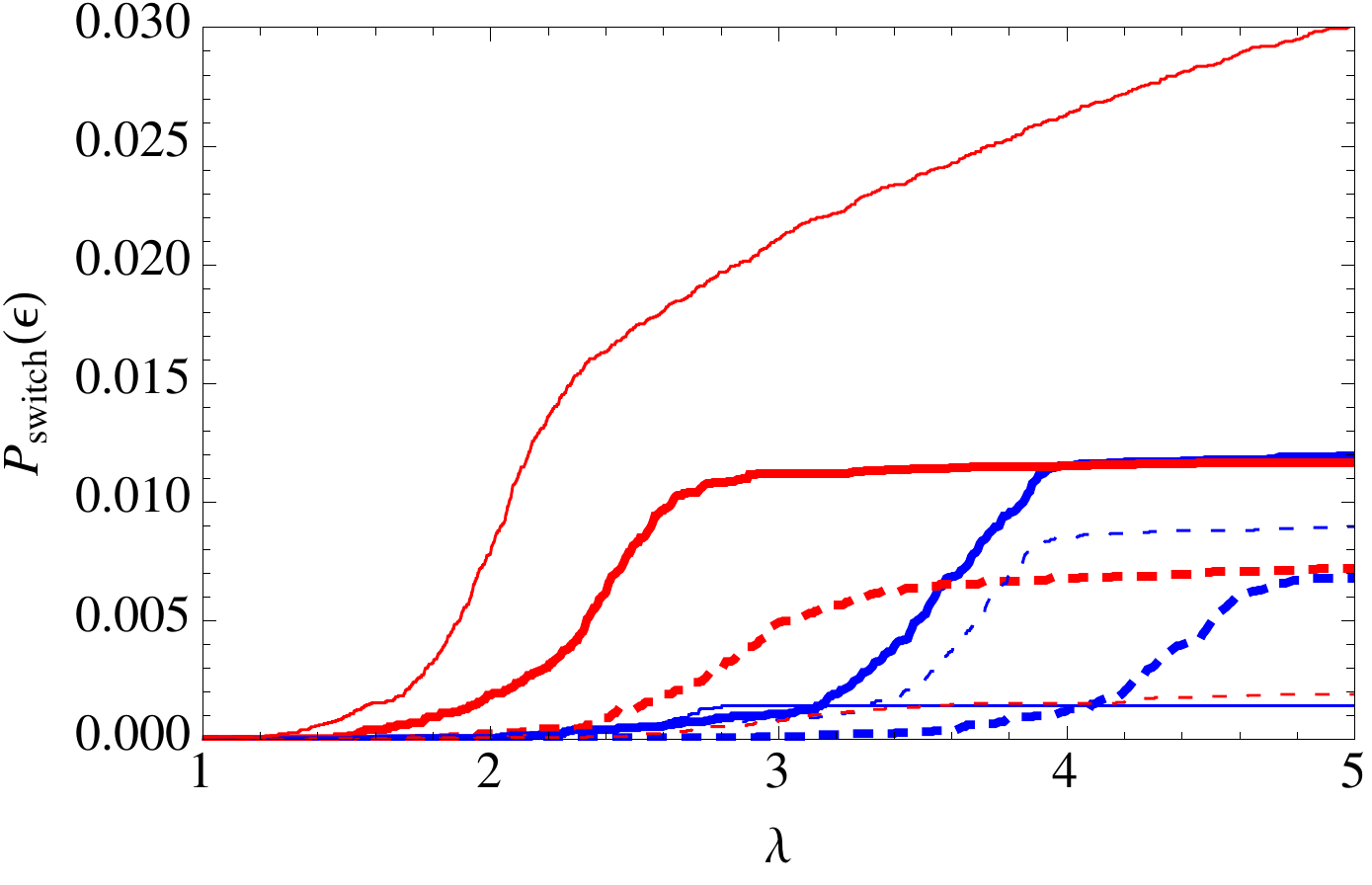}
\caption{Top: SB partner switching probability $P_{switch}$ vs. $g(\lambda)$, for uniaxial stress tests. Bottom: SB partner switching vs $\lambda$ for uniaxial strain. Partner switching is defined as when a sticky bond is broken, and rebonds, but does not rebond to its original partner. Line colors, widths and dashing are the same as in Fig.~\ref{fig:stressstrain2}.}
\label{fig:sbps}
\end{figure}

The top panel of Fig.\ \ref{fig:stressstrain2} shows results for
uniaxial stress. 
At the lower strain rate, which is closer to the quasi static limit\cite{rottler03b}, strong sticky bonds and slow kinetics (blue, dashed curves) yield a
more ``elastic'' response, with a relatively high maximum stress and a relatively low fracture
strain. In contrast, weak sticky bonds and fast
kinetics display opposite trends: much more ``plastic'' behavior, with a lower stress maximum
and higher fracture strain.
Results for uniaxial strain are illustrated in the bottom panel of Fig.\ \ref{fig:stressstrain2}  and show markedly different trends.  
The ``strong, slow'' (Ss4 and Ss55) systems both support a higher
stress at all strains beyond yield, a significantly larger fracture
strain, and clearly larger toughness (Table \ref{tab:WPSPS}) than their counterparts at the
same strain rate.
Note that the uniaxial strain protocol is inherently dilational (system volume $V=\lambda V_0$).  Such a protocol naturally suppresses sticky bond recombination, partner-switching, and self-healing, since SBs are more dispersed.  From this result, we can infer that SB partner switching can play a dominant role in controlling material toughness (cf.\ Fig.\ \ref{fig:sbps}).

\begin{table*}
\caption{Numerical values for the maximum work ($W_{max}$), the fraction of initial SB's surviving ($P_{surv}$) and the fractions of SB's that perform a "partner switch" ($P_{switch}$) for all systems. Values of $W_{max}$ are calculated post-fracture, and the values for $P_{surv}$ and $P_{switch}$ are taken at the strain where the fracture rate $\partial P/\partial\lambda$ is maximized, i.e. at a stretch $\lambda\simeq 4$ ($g(\lambda)\simeq 15$).}
\begin{ruledtabular}
\begin{tabular}{lcccccc}
System & $W_{max} $ & $W_{max}$ & $P_{surv}$ & $P_{surv}$ & $10^{3}P_{switch}$ & $10^{3}P_{switch}$\\
 & [uniaxial stress] & [uniaxial strain] & [uniaxial stress] & [uniaxial strain]& [uniaxial stress] & [uniaxial strain]\\
Wf4 & 202 & 23.3 & 0.962 & 0.980 & $10.4$ & $11.5$\\
Ws4 & 235 & 35.8 &0.979  & 0.984 &$1.67$ & $6.71$\\
Sf4 & 267 & 60.5 & 0.978 & 0.976 & $3.67$ & $11.2$ \\
Ss4 &  295 & 85.4 & 0.989 & 0.988  & $0.667$ & $1.00$ \\
Wf55 & 200 & 14.0 & 0.935 & 0.955 & $16.1$ & $26.0$\\
Ws55 & 228 & 36.4 & 0.977  & 0.972 & $5.04$ & $1.5$\\
Sf55 & 252 & 26.7 & 0.961 & 0.960 &$5.01$ & $1.5$\\
Ss55 & 218 & 56.1 & 0.992 & 0.980 & $1.67$ & $7.1$\\
\end{tabular}
\end{ruledtabular}
\label{tab:WPSPS}
\end{table*}

\begin{figure}
\includegraphics[width=3in]{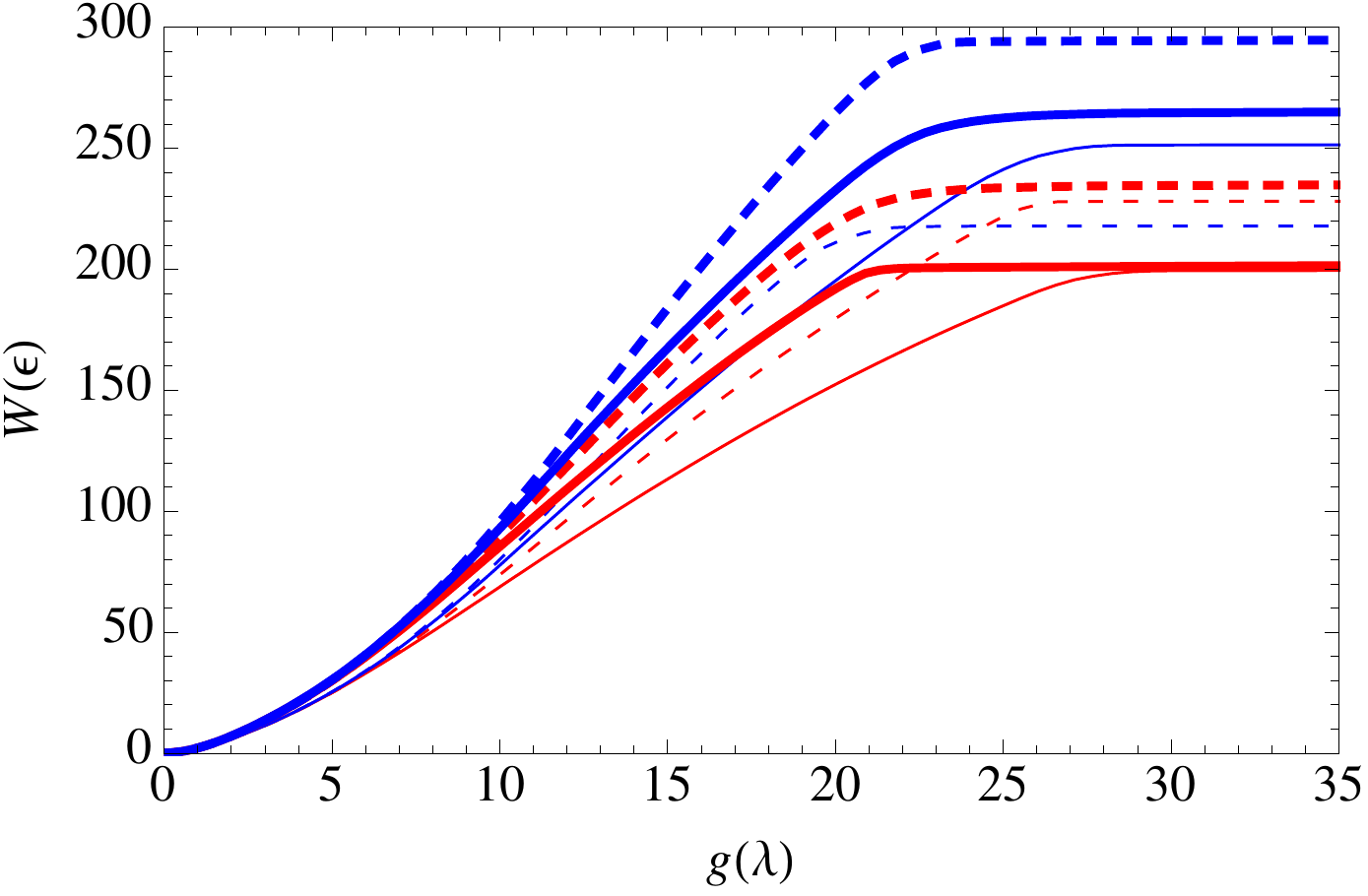}
\includegraphics[width=3in]{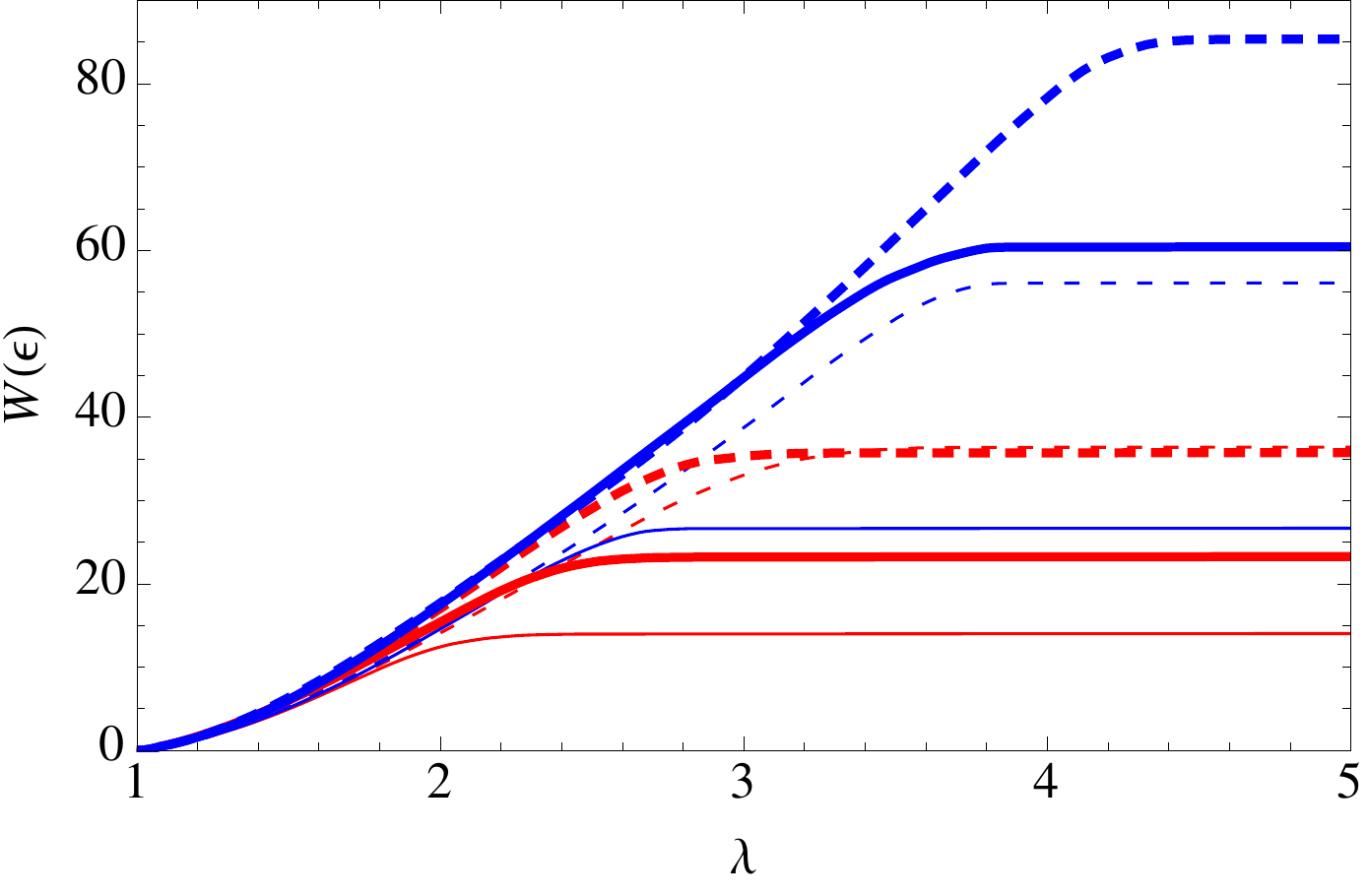}
\caption{Top: Work per unit volume $W(\epsilon)$ vs. $g(\lambda)$ for uniaxial stress tests. Bottom: $W(\epsilon)$ vs $\lambda$ for uniaxial strain.  Line colors, widths and dashing are the same as in Fig.~\ref{fig:stressstrain2}.}
\label{fig:work}
\end{figure}

The data in Figure \ref{fig:stressstrain2} can also be used to isolate
the effect of thermodynamics ($h/k_BT$) from that of chemical
kinetics.  Results for the lower strain rate are as follows: for
``strong'' $h=15$ systems (thin blue curves, Ss55/Sf55), faster kinetics (solid curve,
Sf55) lead to
greater ductility in uniaxial stress, but substantially lower
toughness for uniaxial strain.  The higher maximum stress and lower fracture strain in the top panel indicates that slower kinetics lead to greater elasticity, and in contrast, faster kinetics lead to greater self-healing effects.
For $h=10$ (thin red curves, Ws55/Wf55), results are consistent for both deformation protocols; slower kinetics lead to greater toughness.  
For uniaxial strain (but not stress), it is clear that larger $\tau_{sb}$ always leads to greater toughness (cf.\ Fig. \ref{fig:work}).
One interpretation of this result is that the ``stronger'' and ``slower'' systems are in the limit $\epsilon\tau_{sb} \gg 1$, where fracture is activated by $\lambda$-dependent stretching of the covalent bonds, i.e.\ the sticky bonds behave similarly to (breakable) chemical crosslinks and the overall behavior is rather like that of a standard (non-associating) thermoplastic elastomer.
Note that in all cases studied here, SB-breaking is stress- or strain-activated since $(h/k_BT)$ and quiescent values\cite{hoy09} of $\tau_{sb}$ are very large.

Next we examine the effect of increased strain rate for systems with
the same $h$ and $\tau_{MC}$. For uniaxial stress, for all systems the
peak stress increases and the fracture strain decreases at increased
$\dot{\epsilon}=10^{-4}$ (thick curves).  
The former is expected due to a greater effective viscosity and reduced $\dot{\epsilon}\tau_{sb}$ at the higher strain rates, and the latter is also expected since fracture is stress- or strain-activated.  This competition between higher peak stress and lower fracture strain reveals an interesting competition v/v toughness, which we will examine further below.
Uniaxial strain again shows trends that are different from uniaxial
stress.  For all systems (with the possible exception of weak/slow),
both the peak stress and fracture strain are larger at the higher strain rate.  
The difference may be attributable to the greater chain-stretching at fixed $\lambda$ for uniaxial strain (affine stretching of chains leads to larger extension in uniaxial strain).
Another possible reason for the differences between the deformation protocols is that the cavitation and craze fibril formation\cite{kramer83,rottler03} processes occurring for uniaxial strain have no counterpart in uniaxial-stress deformation.

\begin{figure}
\includegraphics[width=3in]{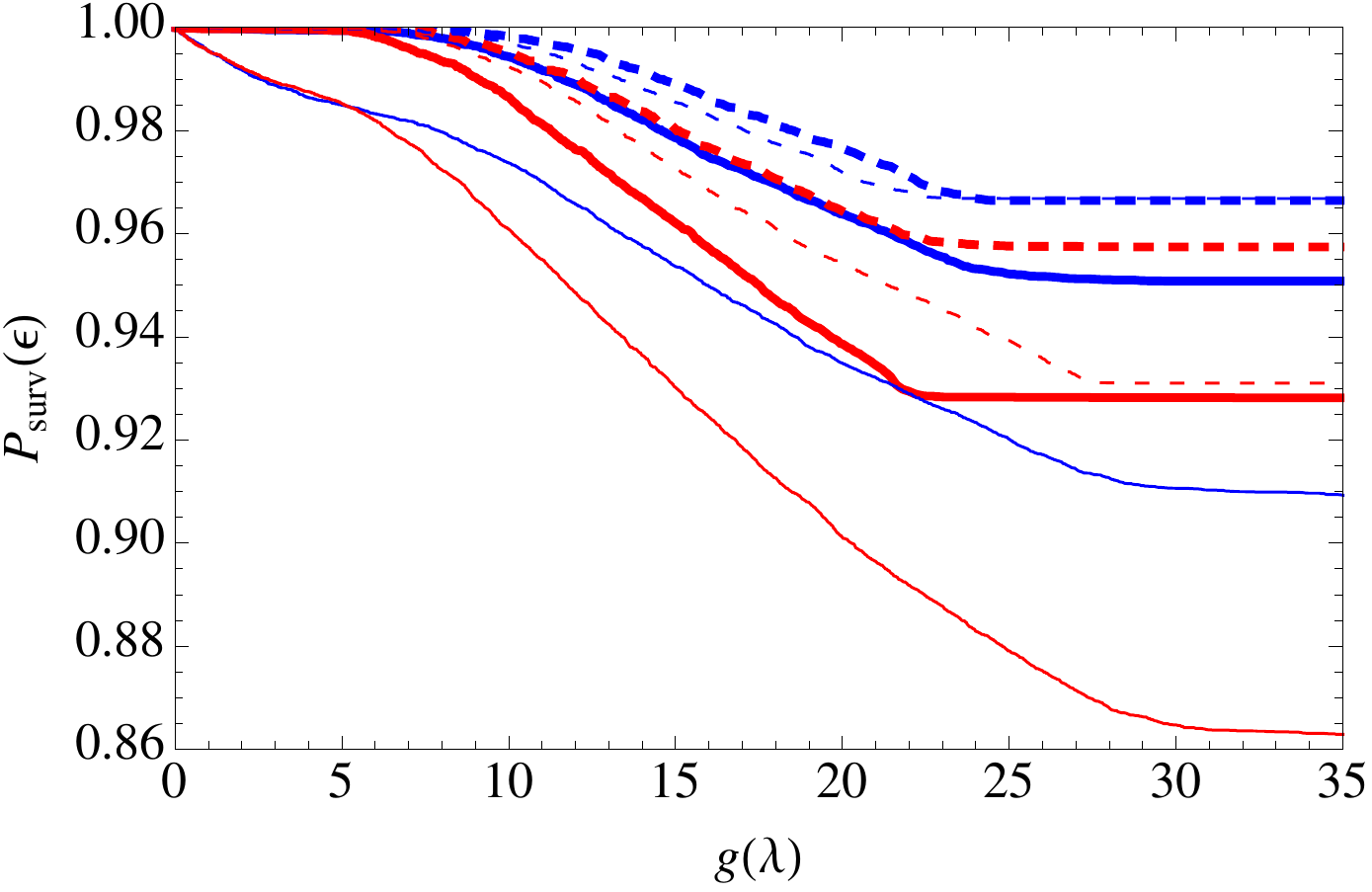}
\includegraphics[width=3in]{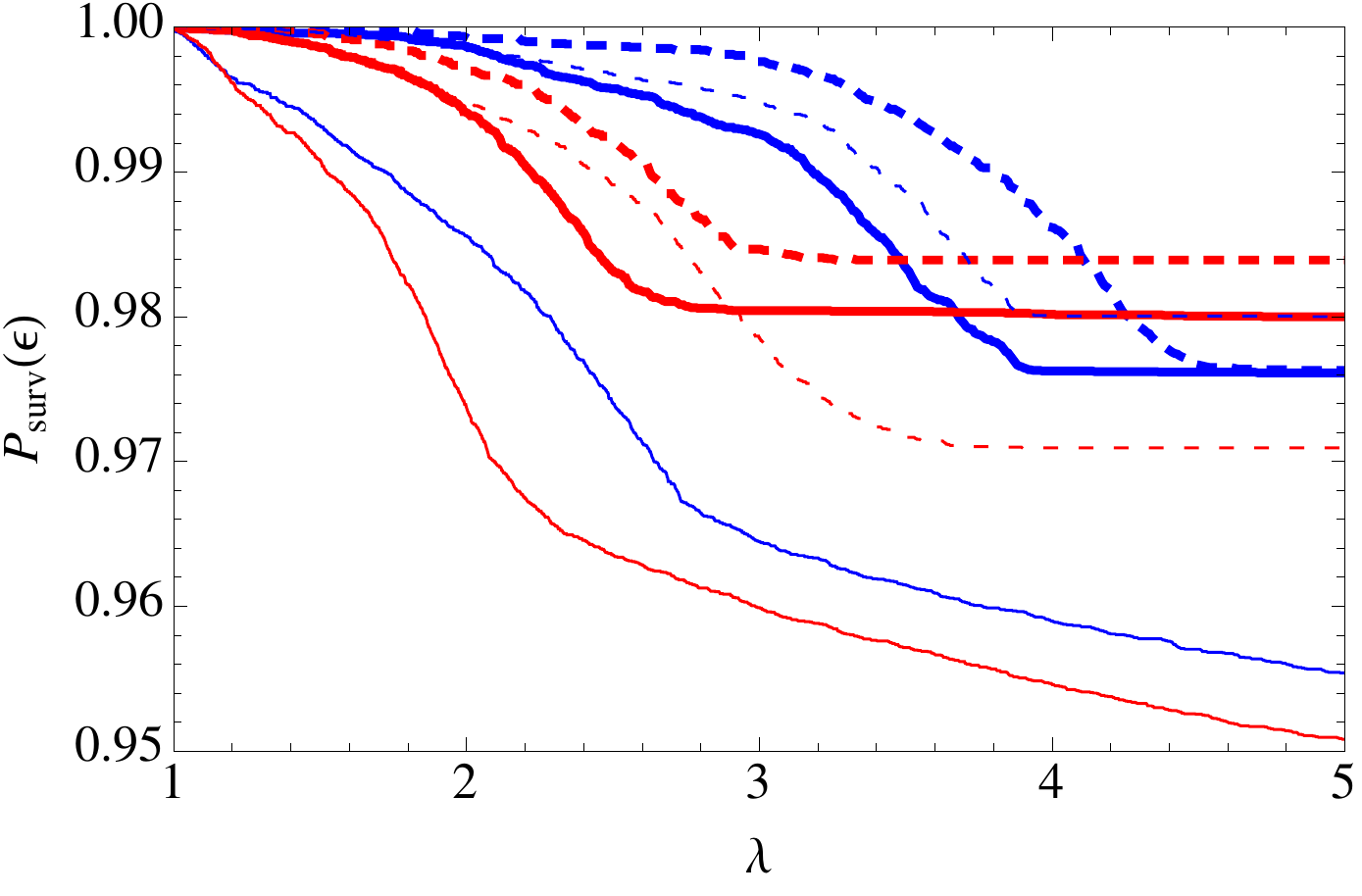}
\caption{Top: Percentage of broken SBs vs g($\lambda$) for uniaxial stress. Bottom: Percentage of broken SBs vs $\lambda$ for uniaxial strain. Here ``breaking'' is defined through single events, i.e. the plot shows the percentage of SB pairs that ``survive'' through the entire deformation up to the given strain.  Line colors, widths and dashing are the same as in Fig.~\ref{fig:stressstrain2}.}
\label{fig:ISB}
\end{figure}

It is interesting that the apparent fracture toughness of these systems is clearly a coupled function of thermodynamics, chemical kinetics, and deformation protocol including strain rate.  
However, only qualitative insights may be obtained by visual inspection of stress strain curves, and analysis is further complicated by the abovementioned competitions. 
We therefore turn to a quantitative, comparative study of toughness (integrated work-to-fracture $W_{max}$) of systems with these different parameters.
Figure \ref{fig:work} plots $W(\epsilon)$ for all systems.
At small strains, for both deformation protocols, work functions lie on the same line of (elastic) response.
As strain increases, work drops below the elastic curves at larger
strains for larger $h$ as well as $\tau_{MC}$, while it decreases for
lower $\dot{\epsilon}$. 
This drop-off coincides with (i.e.\ occurs at the same $\lambda$ as) the onset of sticky bond breaking
(cf.\ Fig. \ref{fig:ISB}).
Finally, at large strains, work plateaus at $W=W_{max}$ (i.e.\ the fracture toughness of systems) as fracture ensues.

Values of $W_{max}$ are ranked in decreasing order from maximum to
minimum as follows: For uniaxial stress: (1) Ss4: (2) Sf4: (3) Sf55:
(4) Ws4: (5) Ws55: (6) Ss55: (7) Wf4: (8) Wf55. 
The work to fracture varies by a factor of $\sim 1.5$ from maximum to minimum.
As expected and shown in Table \ref{tab:WPSPS}, the toughest systems are the ``strong, slow'' systems
deformed at the higher strain rate (Ss4) and the most brittle systems
are the ``weak, fast'' systems deformed at the lower strain rate (Wf55).
However, these dependencies are not monotonic; systems with weaker SBs under high-strain-rate conditions can be tougher than systems with stronger SBs deformed at low strain rates, illustrating the complexity of these systems' mechanical response.
As we have surmised above, the fracture toughness shows a deformation protocol dependence. 
The ordering of $W_{max}$ from maximum to minimum is different from that for uniaxial stress: (1) Ss4: (2) Sf4: (3) Ss55: (4) Ws55: (5) Ws4: (6) Wf55: (6) Ss55: (7) Wf4: (8) Wf55. 
Further, in contrast to the small fractional variations in $W_{max}$
for uniaxial stress, for uniaxial strain the toughest systems'
$W_{max}$ is about six times higher (Table \ref{tab:WPSPS}) than that of the most brittle
systems.
We expect this more dramatic difference is closely associated with the dilatative nature of crazing.
Specifically, the toughest (Ss and Sf) systems (similarly to entangled polymer
glasses\cite{kramer83,rottler03}) show strain hardening at the end of
the craze drawing plateau, indicating chains are stretching between SB
junctions, while the more brittle systems fracture before the strain hardening regime is reached.

\begin{figure}
\centering
\begin{subfigure}
  \centering
\includegraphics[width=1.625in]{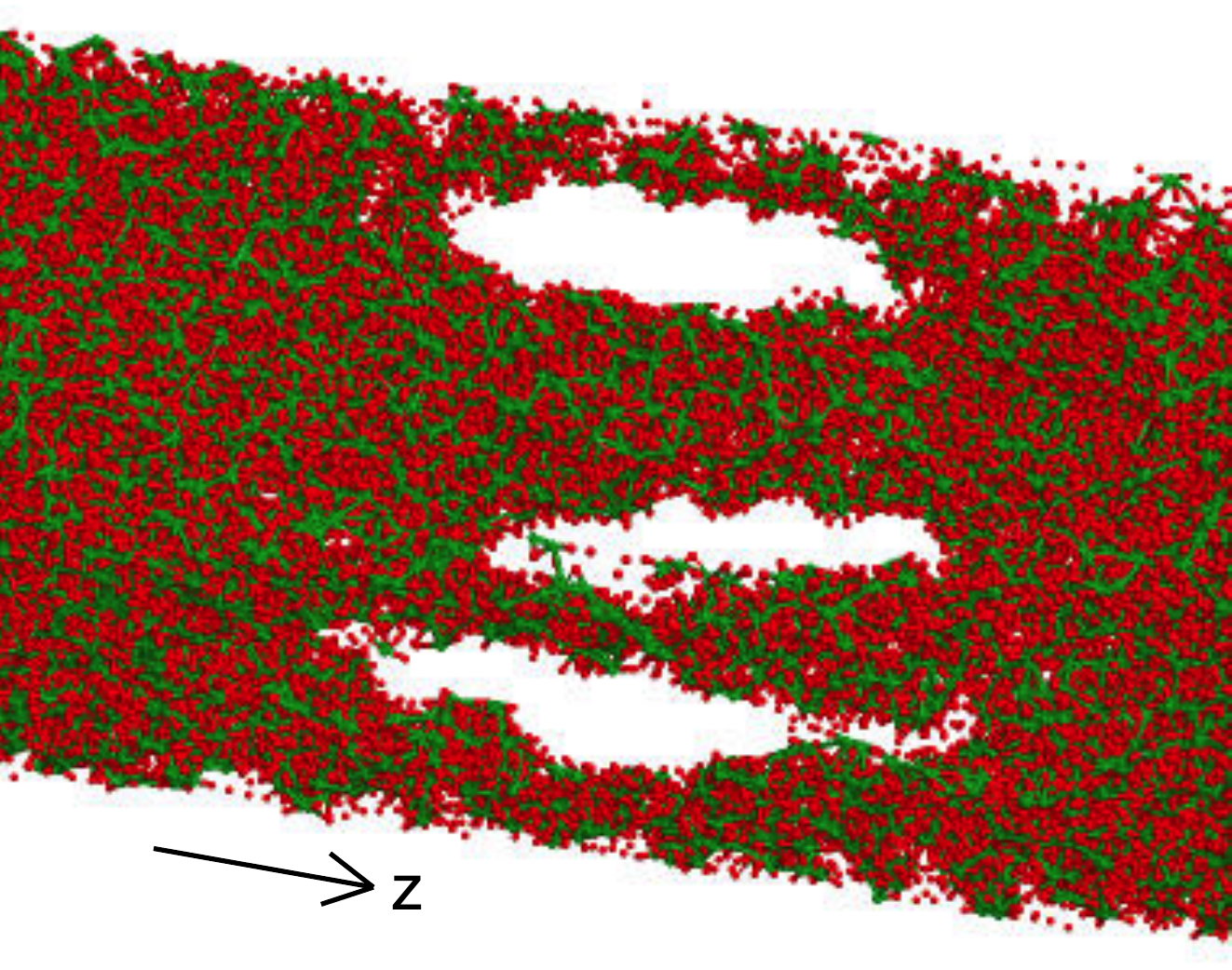}
\end{subfigure}
\begin{subfigure}
  \centering
\includegraphics[width=1.625in]{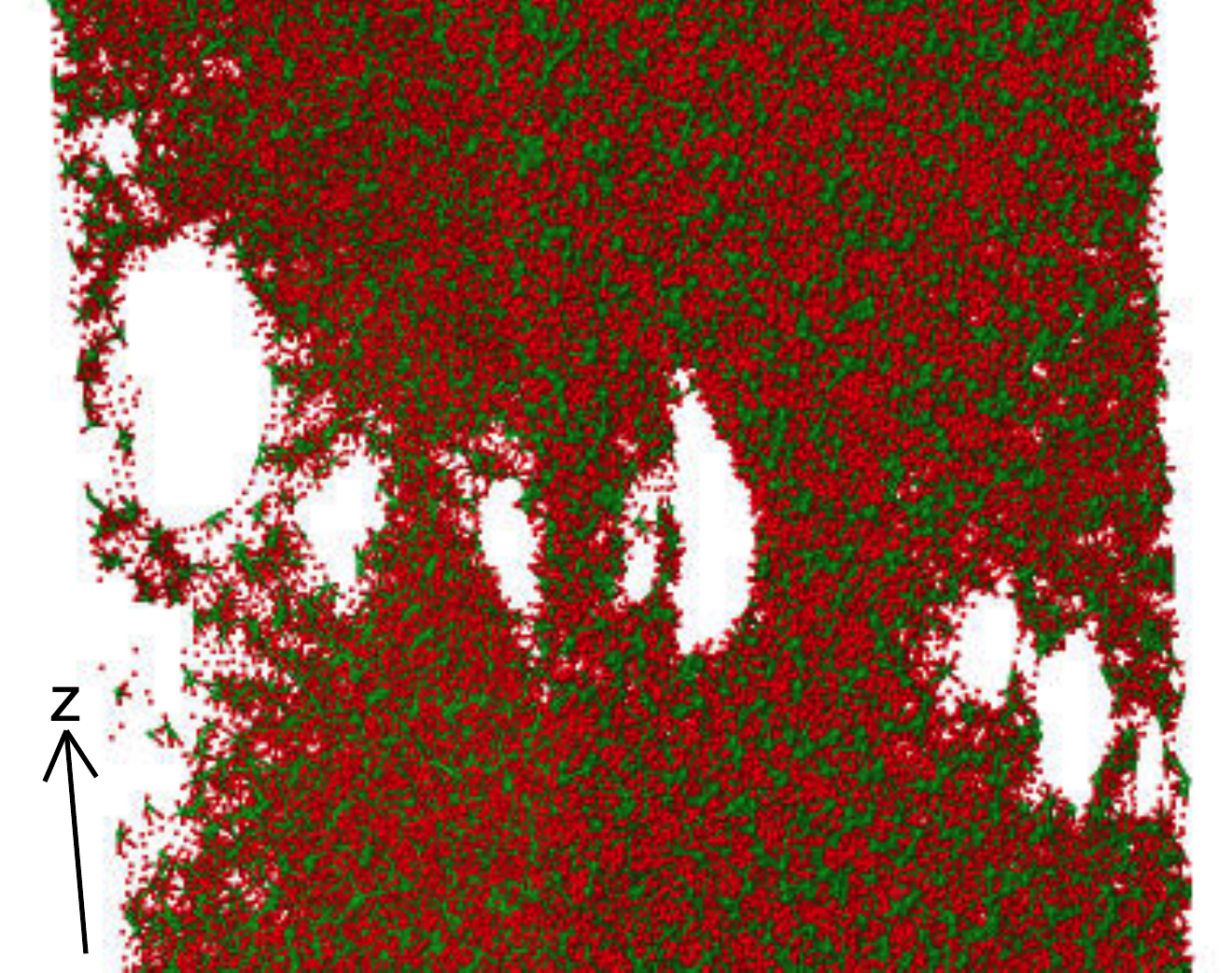}
\end{subfigure}
\caption{Relating $P_{surv}$ to fracture geometry.  Left: The systems with the highest fraction of surviving bonds fracture along a single plane via formation of a single large void.  Right: The systems with the lowest fraction of surviving bonds exhibit a more complex fracture geometry.  Both images show only small cross-sections of systems; regions which are not shown remain in a dense, nearly undisturbed state.}
\label{fig:movies}
\end{figure}

To relate these differences to the underlying associating polymer physics, we  next examine measures of sticky bond breaking and partner-switching.
Figure \ref{fig:ISB} illustrates the breaking of sticky bonds,
i.e.\ the strain-dependent SB survival probability $P_{surv}$.  The
onset of SB-breaking closely corresponds to the divergence of
stress-strain curves for different systems that occurs beyond the
elastic limit (Fig.~\ref{fig:stressstrain2}).
This generally occurs at lower strains for smaller $h$, faster kinetics, and lower strain rates.
Differences can be quite dramatic, e.g.\ the differences between
uniaxially stressed $h=10$ systems at low and high strain rates are
large because: \textbf{(i):} more slowly deformed systems (*55) are in a regime where $\dot{\epsilon}\tau_{sb} < 1$: \textbf{(ii):} more slowly deformed systems are more ductile. 
In all cases, the SB-breaking rate accelerates to a maximum (indicated by the maximum slopes in Fig.~\ref{fig:ISB}) as fracture initiates, then plateaus after fracture as the systems are no longer under stress.  
One interesting if unsurprising effect is that $P_{surv}$ at the termination of deformation can be much lower for uniaxial stress.  
This is because fracture under uniaxial stress can be much less localized than crazing-type fracture.
Visual inspection (Figure \ref{fig:movies}) shows that systems with high $P_{surv}$ fracture along a single plane with a single large void, while systems with lower $P_{surv}$ form many smaller voids and fracture in multiple locations simultaneously.

Next we turn to other measures of stress relaxation.
Sticky-bond recombination, wherein a bond pair $A-B$ (where $A$ and
$B$ are distinct sticky monomers) breaks and then reforms, is a fast
stress-relaxation mechanism.  While the bond is ``open'', stress can
locally relax, and the stress previously borne by the open bond is
transferred to other closed bonds.  This eases the onset of the
mechanical instability corresponding to fracture.
We find that SB recombination is
important only for the lower strain rate and in systems with fast
kinetics in the uniaxial stress protocol (thin, solid curves, see Fig.~\ref{fig:sbps}).  These trends are expected since recombination requires a finite amount of time (i.e.\ of order $\tau_{MC}$), and higher strain rates and slower kinetics both cause broken $A-B$ pairs to move further away from one another before they can recombine.
Far more significant from the point of view of stress relaxation in these systems is sticky-bond partner exchange, wherein an $A-B$ pair breaks and $A$ recombines with a different sticky monomer $C$.  This is typically an irreversible process corresponding to plastic deformation.
As shown in Figure \ref{fig:sbps}, the onset of recombination
corresponds to the onset of bond-breaking.  The slope is similarly
maximized when systems are plastically deforming (Table \ref{tab:WPSPS}).
SB partner switching is most
significant for ``weak, fast'' bonds and at low strain rates (Wf55).  As shown in Fig.\ \ref{fig:stressstrain2}, these systems also display the greatest degree of plastic flow, i.e.\ flow at nearly constant stress.  
We expect that SB partner switching is the primary mechanism of self-healing in these materials.
In other words, partner switching allows for greater ductility because it allows the materials to heal themselves even when a finite strain rate is applied.

\section{Conclusions}

The properties of thermoreversibly associating polymer melts and glasses have attracted great interest over the past fifteen years on a variety of fronts.
Of particular promise is the ability to control the thermodynamics and
kinetics of the sticky bonds independently, e.g.~by varying the
chemistry of metal-ligand groups forming the bonds\cite{yount03,yount05,loveless05,Rinderspacher201296,doi:10.1021/ma401077d}.
Specifically, if $h$ is the binding energy of sticky monomers, $T$ is temperature,
$\nu_0$ is a characteristic kinetic ``attempt rate'' determined by the
geometry of the binding groups, and $\tau_{sb}$ is the lifetime of
sticky bonds, then $\tau_{sb} = f(h/k_BT)/\nu_0$.  
In other words, $\tau_{sb}$ can be written as a product of factors
controlled by thermodynamics (i.e.~$f(h/k_BT)$) and chemical
kinetics.  
Frequency-dependent properties such as the shear modulus $G(\omega)$
can often be scaled by these kinetic rates, i.e. systems with
different $G(\omega)$ have the same $G(\omega/\nu_0)$.\cite{yount03,yount05,loveless05}  
 However, this collapse can break down when $\nu_0$ is comparable to the characteristic relaxation rates $\tau_{pol}^{-1}$ of the parent chains in the absence of sticky bonding, in a complex regime of interplay between SB and parent-chain relaxation dynamics.\cite{hoy09}
It is exactly this complex regime that we have studied here, with emphasis on the nonlinear mechanics of glassy systems.

We have characterized the mechanical properties of model thermoreversibly associating dendrimer glasses
using a hybrid molecular dynamics/Monte Carlo method.\cite{hoy09}  
The short, unentangled ``Y'' architecture  employed here is a simple model of a dendrimer with unentangled arms; the sticky bond network produces an entangled-like mechanical response that is not (to first order) architecture-specific.
Such glasses are of interest because they are easily melt-processable yet can show remarkably high ductility and fracture toughness.  
We examined the entire range of mechanical response from the elastic regime through fracture.
At small strains, all systems fall on a common stress-strain curve since SB-breaking has not yet initiated.  Local ``yield''  (i.e.\ the divergence of stress curves for systems with different SB parameters) corresponds to the onset of SB-breaking and naturally occurs in systems with thermodynamically weak SBs possessing fast chemical kinetics.  
All systems exhibit entangled-like response\cite{haward97,rottler03} at small and moderate strains because the sticky bonds act like (transient) chemical crosslinks.
For larger strains, mechanical strength is maximized by ``strong, slow'' (Ss) sticky bonds, while ductility is maximized by ``weak, fast''  (Wf) SBs because they ``partner-switch'' in a dynamical self-healing process.
Investigations of such healing processes during active deformation are in their infancy; to our knowledge, this is the first time they have been reported in thermoreversibly associating glasses.

We have also examined toughness (work-to-fracture).  In general, toughness is maximized for Ss systems and minimized for Wf systems.  
However, this result is nontrivial since Wf systems exhibit larger fracture strains under some deformation conditions.
Strain-rate dependence studies show that these systems exhibit a complex rheology; our systems are often, but not always, tougher at higher strain rates, but they are more ductile at lower strain rates.
This illustrates the need for further studies over a broad range of strain rate to examine the response under conditions ranging from quasi-static to shock loading.
Future work will consider polymers of different (e.g.\ physically entangled) topology, varying SB concentration, response under other deformation protocols such as creep, and temperature dependence.

\section{Acknowlegements}

All simulations were performed using an enhanced version of LAMMPS.\cite{plimpton95}  
This work was partially funded by ARL contract
TCN-11042 and the US ARL Enterprise for Multi-Scale Research of Materials. AS was supported by the REU program at USF (NSF Grant No.\ DMR-1263066). We would like to thank Yelena R. Sliozberg and Robert H. Lambeth for helpful discussions.


\end{document}